\documentclass[conference]{IEEEtran}
\IEEEoverridecommandlockouts

\usepackage{array,multirow}
\usepackage{booktabs} 
\usepackage{graphicx}
\usepackage{caption}
\usepackage{subcaption}
\usepackage{amssymb}
\usepackage{bbm}
\usepackage{threeparttable}
\usepackage{booktabs}
\usepackage{cite}
\usepackage{amsmath,amssymb,amsfonts}
\usepackage{algorithmic}
\usepackage{graphicx}
\usepackage{mathtools}
\usepackage{textcomp}
\usepackage{xcolor}
\usepackage[utf8]{inputenc}
\usepackage{pgfplots}
\DeclareUnicodeCharacter{2212}{−}
\usepgfplotslibrary{groupplots,dateplot}
\usetikzlibrary{patterns,shapes.arrows}
\pgfplotsset{compat=newest}

\def \fwidth{0.99\columnwidth}
\def \fheight {0.6\columnwidth}

\def\BibTeX{{\rm B\kern-.05em{\sc i\kern-.025em b}\kern-.08em
    T\kern-.1667em\lower.7ex\hbox{E}\kern-.125emX}}

\usepackage[acronyms,nonumberlist,nopostdot,nomain,nogroupskip]{glossaries}

\newacronym{sla}{SLA}{Service Level Agreement}
\newacronym{qos}{QoS}{Quality of Service}
\newacronym{qoe}{QoE}{Quality of Experience}
\newacronym{embb}{eMBB}{enhanced Mobile BroadBand}
\newacronym{urllc}{URLLC}{Ultra Reliable Low Latency Communication}
\newacronym{mmtc}{mMTC}{massive Machine Type Communication}
\newacronym{sdn}{SDN}{Software Defined Networking}
\newacronym{nfv}{NFV}{Network Function Virtualization}
\newacronym{ns}{NS}{Network Slicing}
\newacronym{iot}{IoT}{Internet of Things}
\newacronym{bbu}{BBU}{BaseBand Unit}
\newacronym{psc}{PSC}{Public Safety Communication}
\newacronym{lmr}{LMR}{Land Mobile Radio}
\newacronym{bb}{BB}{Broadband}
\newacronym{ps}{PS}{Public Safety}
\newacronym{uav}{UAV}{Unmaned Aerial Vehicle}
\newacronym{bs}{BS}{Base Station}
\newacronym{dqn}{DQN}{Deep Q-Network}

\newacronym{marl}{MARL}{Multi-Agent Reinforcement Learning}
\newacronym{drl}{DRL}{Deep Reinforcement Learning}
\newacronym{ml}{ML}{Machine Learning}
\newacronym{rl}{RL}{Reinforcement Learning}
\newacronym{pomdp}{POMDP}{Partially Observable Markov Decision Process}
\newacronym{mdp}{MDP}{Markov Decision Process}
\newacronym{cnn}{CNN}{Convolutional Neural Network}
\newacronym{nn}{NN}{Neural Network}
\newacronym{pdf}{PDF}{Probability Density Function}
\newacronym{radam}{RAdam}{Rectified Adam}
\newacronym{cdf}{CDF}{Cumulative Distribution Function}
\newacronym{relu}{ReLU}{Rectified Linear Unit}
\newacronym{adam}{Adam}{Adaptive moment estimator}
\newacronym{ac}{AC}{Actor Critic}
\newacronym{a2c}{A2C}{Advance Actor Critic}
\newacronym{gnn}{GNN}{Graph Neural Network}
\newacronym{fl}{FL}{Federated Learning}
\newacronym{sarsa}{SARSA}{State-Action-Reward-State-Action}
\newacronym{mc}{MC}{Markov Chain}
\newacronym{mec}{MEC}{Mobile Edge Computing}

\newacronym{fifo}{FIFO}{First In First Out}
\newacronym{cvo}{CVO}{Critical Voice}
\newacronym{cvi}{CVI}{Critical Video}
\newacronym{ncvo}{NCVO}{Non-Critical Voice}
\newacronym{ncvi}{NCVI}{Non-Critical Video}

\makeatletter
\def\endthebibliography{%
	\def\@noitemerr{\@latex@warning{Empty `thebibliography' environment}}%
	\endlist
}
\makeatother

\definecolor{color0}{HTML}{00008c}
\definecolor{color1}{HTML}{5c0093}
\definecolor{color2}{HTML}{940095}
\definecolor{color3}{HTML}{c3208d}
\definecolor{color4}{HTML}{ec4373}
\definecolor{color5}{HTML}{ff7600}

\DeclareMathOperator*{\argmax}{arg\,max}

\IEEEoverridecommandlockouts
\newcommand\copyrightnotice{%
	\begin{tikzpicture}[remember picture,overlay]
	\node[anchor=north,yshift=-10pt] at (current page.north) {\fbox{\parbox{\dimexpr\textwidth-\fboxsep-\fboxrule\relax}{\footnotesize \textcopyright 2021 IEEE. This paper has been submitted to IEEE ICC 2022. Personal use of this material is permitted. Permission from IEEE must be obtained for all other uses, in any current or future media, including reprinting/republishing this material for advertising or promotional purposes, creating new collective works, for resale or redistribution to servers or lists,	or reuse of any copyrighted component of this work in other works.}}};
	\end{tikzpicture}
}

\begin{document}

\title{\vspace{.5cm} No Free Lunch: Balancing Learning and Exploitation at the Network Edge}

\author{\IEEEauthorblockN{Federico Mason}
\IEEEauthorblockA{\emph{Dept. of Information Engineering} \\
\emph{University of Padua}\\
Via Gradenigo 6/b, Padua, Italy \\
masonfed@dei.unipd.it}
\and
\IEEEauthorblockN{Federico Chiariotti}
\IEEEauthorblockA{\emph{Dept. of Electronic Systems} \\
\emph{Aalborg University}\\
Fredrik Bajers Vej 7C, Aalborg, Denmark \\
fchi@es.aau.dk}
\and
\IEEEauthorblockN{Andrea Zanella}
\IEEEauthorblockA{\emph{Dept. of Information Engineering} \\
\emph{University of Padua}\\
Via Gradenigo 6/b, Padua, Italy \\
zanella@dei.unipd.it}
\thanks{This work was supported by Consortium GARR through the “Orio Carlini” scholarship 2019, and partly by the Danmarks Frie Forskningsfond (DFF) through the WATER project.}
}
\maketitle

\copyrightnotice

\begin{abstract}
Over the last few years, the \gls{drl} paradigm has been widely adopted for 5G and beyond network optimization because of its extreme adaptability to many different scenarios.
However, collecting and processing learning data entail a significant cost in terms of communication and computational resources, which is often disregarded in the networking literature.
In this work, we analyze the cost of learning in a resource-constrained system, defining an optimization problem in which training a \gls{drl} agent makes it possible to improve the resource allocation strategy but also reduces the number of available resources.
Our simulation results show that the cost of learning can be critical when evaluating \gls{drl} schemes on the network edge and that assuming a cost-free learning model can lead to significantly overestimating performance.
\end{abstract}

\begin{IEEEkeywords}
Reinforcement Learning, Continual Learning, Network Slicing, Mobile Edge Computing
\end{IEEEkeywords}

\glsresetall

\section{Introduction}
The orchestration of next-generation mobile networks is beyond the capabilities of human-designed algorithms, as it is characterized by multiple objectives and fast dynamics, with several classes of traffic having highly specific activity patterns and \gls{qos} guarantees~\cite{navarro2020}.
In this context, machine learning is essential to allow the network protocols to dynamically adapt to different scenarios without the need to manually reconfigure the entire system~\cite{morocho2019machine}.
In particular, the combination of \gls{rl} principles with deep learning, also known as \gls{drl}~\cite{mnih2015human}, is one of the most promising tools for the optimization of 5G and beyond networks~\cite{xiong2019deep}.

The training of \gls{drl} models in complex environments is still very computationally expensive~\cite{gomez2020exploring}, and cannot be always performed in advance because of the rapid changes that characterize future mobile networks.
Therefore, it is fundamental for 5G and beyond systems to support online training on the network edge, exploiting a continual learning approach~\cite{zenke2017continual}.
In this context, the training updates are either performed directly on the edge nodes, according to the \gls{mec} paradigm, or they are offloaded to more powerful Cloud servers, reducing the consumption of local resources but requiring the transmission of large amounts of data.

In a \gls{mec} scenario, the training cost creates a fundamental trade-off: updating a \gls{drl} model with new experience can improve the learned policy, increasing its efficiency in the target task, but also subtracts some resources from that very same task.
Disregarding this trade-off and assuming that training is a \emph{free action} may have serious consequences, degrading the performance of the system that should be optimized~\cite{villacca2021online}.
The machine learning research community is starting to become aware of this issue~\cite{thompson2020computational}, focusing on model compression and lightweight learning techniques to reduce the burden on the edge hardware~\cite{liu2020bringing} and considering the \emph{cost of learning} in the design of neural networks meant to operate on resource-constrained devices~\cite{chan2018learning}.
In particular, the \gls{fl} approach can reduce the computational load of learning systems by performing the training of the target algorithms in a distributed fashion~\cite{khan2020federated}. 
In the past years, many frameworks to reduce the computation and communication cost of \gls{fl} have been proposed: for a deeper review on these topics, we refer the reader to~\cite{chen2019deep, luo2021cost}.

To the best of our knowledge, the resource efficiency of \gls{drl} techniques in future network scenarios is a relatively unexplored topic~\cite{jang2020knowledge}, as most researches in the literature still neglect the cost of the training, separating the learning process from the optimization even in online applications.
In this work, we attempt to model the cost of learning explicitly, defining an optimization problem that balances learning and system optimization.
The objective is to identify learning strategies that maximize the system performance during training, accounting for the cost of the learning itself.
Our results show that adapting the number of training updates is a key factor for the optimization of \gls{mec} systems, while considering an ideal case with free learning actions may lead to a significant overestimating of the real performance.

The rest of this paper is organized as follows: Sec.~\ref{sec:meta-model} presents the cost of learning problem, which is then applied in a network slicing use case in Sec.~\ref{sec:system-model}. The results of our online learning are presented in Sec.~\ref{sec:results}, and Sec.~\ref{sec:concl} concludes the paper and presents some possible avenues of future work.

\section{Cost of Learning Model}
\label{sec:meta-model}

We consider a typical network management application, in which a \gls{drl} agent tries to maximize system performance by allocating communication or computational resources to different users. The environment is modeled as a \gls{mdp} defined by the 4-tuple $(\mathcal{S}, \mathcal{A}, P, R)$, where $\mathcal{S}$ is the state space, $\mathcal{A}$ is the action space, $P:\mathcal{S}\times\mathcal{A}\rightarrow\mathcal{S}$ is the state transition probability function and $R:\mathcal{S}^2\times\mathcal{A}\rightarrow\mathbb{R}$ is the reward function. 
Hence, time is discretized into slots $t=0,1,2,...$ and, at any slot, the agent chooses a new action according to a policy $\pi:\mathcal{S} \rightarrow \mathcal{A} $.
The function $P_{a_t}(s_{t}, s_{t+1})$ is the probability that action $a_t$ in state $s_{t}$ at time $t$ will lead to state $s_{t+1}$ at time $t+1$, while $R_{a_t}(s_{t}, s_{t+1})$ is the immediate reward received after a transition from $s_{t}$ to $s_{t+1}$ due to action $a_t$. 
The goal of the agent is to learn the optimal policy $\pi^*$, which maximizes the expected discounted return $G(t)$:
\begin{equation}
G(t) = \mathbb{E}\left[ \sum_{\tau=t}^{\infty} \gamma^{\tau-t} R_{a_{\tau}}(s_\tau, s_{\tau+1})\ \middle|\ \pi,s_t \right],\label{eq:long_term}
\end{equation}
where $\gamma \in [0,1)$ is the so-called \emph{discount factor}. 

We assume that the environment dynamics change over time, which makes it impossible to explore the environment offline, i.e., collect data with a pre-determined policy and perform training on this static dataset before the agent deployment. 
More specifically, we consider that the environment is organized into episodes $k=0,1,...$, and that each episode lasts $T$ slots.
Hence, we assume that $P(\cdot)$ and $R(\cdot)$ are stochastic processes that change over time in a step-wise manner, after a \emph{coherence period} of $K$ episodes. 
In general, we can assume that the agent is aware of the environment evolution and that it is reinitialized at the end of every coherence period.
Accelerating the training might be possible by exploiting \emph{transfer learning}, using previous experience to avoid periodically starting a new learning process from scratch~\cite{tan2018survey}. However, the implementation of such techniques is out of the scope of this work.

We define the expected reward for an episode as
\begin{equation}
	\overline{R}(k) = \mathbb{E}\left[\sum_{t=0}^{T-1}  \frac{R_{a_t} (s_t, s_{t+1})}{T}\right],
\end{equation}
and we set the goal of maximizing the expected reward $\overline{R}_K= \sum_{k=0}^{K-1}\overline{R}(k)$ during the $K$ episodes constituting a single coherence period.
In general, this is achieved by making the agent policy converge to the optimum in the shortest time possible.
However, in our scenario, training the agent consumes some of the resources (computational or communication, depending on the architecture) that should be assigned to the users.
Hence, each learning update has a direct impact on the system performance, and there is a trade-off between convergence speed and cost of learning.

To model this aspect, we assume that each episode is divided into two subsequent phases: first, an \emph{exploitation} phase, in which the learned strategy is applied, then an \emph{update} phase devoted to the agent training.
In particular, the two phases last  $T_{\pi}$ and  $T_{\rho} = T -  T_{\pi}$ slots, respectively, so that the agent's actions are determined by the learned policy $\pi$ during the first $T_{\pi}$ slots of any episode $k$, while the agent follows a pre-determined policy $\rho$ during the last $T_{\rho}$ slots, i.e.,
\begin{equation}
a_t =
\begin{cases}
\pi(s_t), \quad & t \in\{ 0, 1,..., T_{\pi}-1\}; \\
\rho(s_t), \quad & t\in\{T_{\pi}, T_{\pi} + 1,..., T - 1\},
\end{cases}
\end{equation}
where $t$ is the slot index within the same episode. 

The policy $\rho$ ensures that all or part of the system resources are used for the agent training, making it possible to update the policy $\pi$ with the new experience that the agent has gained.
On the other hand, since $\rho$ subtracts some of the resources from the users, this strategy leads to sub-optimal performance, decreasing the reward collected during the episode. 

From a practical perspective, the value of $T_{\rho}$ determines the amount of time and resources devoted to the learning process. As $T_{\rho}$ increases, so does the number of experience samples used to train the agent after each episode.
This makes it possible to accelerate the convergence speed of the agent, thus improving the immediate reward gained during the exploitation phases of the next episodes.
On the other hand, increasing $T_{\rho}$ also shortens those exploitation phases, because more time is spent in the update phases in which the sub-optimal policy $\rho$ is used.
In particular, the average reward during the $k$-th episode, considering a fixed $T_{\rho}$, is given by:
\begin{equation}
 \overline{R}(k|T_{\rho})=\left(\frac{T-T_{\rho}}{T} \overline{R}_{\pi_k}(k) + \frac{T_{\rho}}{T} \overline{R}_{\rho}(k) \right),
\end{equation}
where $\pi_k$ is the policy learned by the agent in the $k$-th episode, while $\overline{R}_{\pi_k}(k) = \mathbb{E}[\overline{R}(k);\pi_k]$ and $\overline{R}_{\rho}(k) = \mathbb{E}[\overline{R}(k);\rho]$ are the expected average rewards gained using policies $\pi_k$ and $\rho$, respectively. 
The optimal value of $T_p$ depends on multiple factors, including the coherence time of the underlying non-stationary \gls{mdp}.

In this work, we consider two possible approaches for balancing exploitation and training.
In the first, we adopt the naive assumption that the amount of resources assigned to the learning task is constant over time, and we analyze how $T_{\rho}$ affects the system performance.
In the second approach, instead, we assume that the amount of training resources can be adapted over time, based on the convergence speed of the learning agent, and we study the trade-off between convergence speed and effectiveness of the learned strategy.
In the following, we formally define the optimization problems underlying the two approaches, which will be successively analyzed and compared in Sec.~\ref{sec:results}, considering a \gls{mec} system as a use-case scenario. 

\subsection{Constant update duration}\label{ssec:const}

Given the coherence period duration $K$, the first optimization problem aims at determining the optimal $T_{\rho}$ to maximize $\overline{R}_K$, while assuming that $T_{\rho}$ is constant over time:
\begin{align}
T_{\rho}^*=\argmax_{T_{\rho}\in\{0,\ldots,T\}}\left( \sum_{k=0}^{K-1} \frac{\overline{R}(k|T_{\rho})}{K}\right).
\end{align}

\subsection{Adaptive update duration}
\label{ssec:adaptive}

The second approach assumes that the agent can determine when the policy $\pi$ converges to the optimal one. 
In particular, after discovering the optimal policy, the agent sets $T_{\rho}$ to zero, fully exploiting the system resources.
If we define the number of episodes until convergence as $\eta(T_{\rho})$, we have the following optimization problem:
\begin{equation}
T_{\rho}^*=\argmax_{T_{\rho}\in\{0,\ldots,T\}}\left(\sum_{k=0}^{\eta(T_{\rho})-1} \frac{\overline{R}(k|T_{\rho})}{K}+\left(1-\frac{\eta(T_{\rho})}{K}\right)\overline{R}_{\pi^*}\right).
\end{equation}
In this case, we expect $T_{\rho}^*$ to be higher, since a quicker convergence to the optimal policy allows the system to terminate the training process and fully dedicate its resources to the users.

\section{Use-case}
\label{sec:system-model}

To test the benefits of our cost-aware learning framework, we consider a \gls{ns} scenario, where a set of users with heterogeneous requirements transmits data through the uplink.
We assume that the traffic is divided into \emph{slices} that depend on the applications' \gls{qos} requirements, and the \gls{bs} needs to allocate the capacity of a backhaul link to guarantee the best possible performance for each slice.
The link resources are managed by a \gls{drl} agent, whose actions depend on the state of the slice buffers at the \gls{bs}. 

%We now define the system model for our use case, which represents a slicing scenario in a 5G network with heterogeneous traffic. In our scenario, the bottleneck of the system is the backhaul from the \gls{bs} to the core network, which needs to support both the uplink traffic and the transmission of experience samples from the \gls{drl} agent that controls the \gls{bs} to the Cloud server to which the learning updates are offloaded.

\subsection{Communication model}
\label{sec:comm-model}

We assume that time is discretized into slots $t = 0,1,2,...$ of length $\tau$ and that, in each slot $t$, each user $u \in \mathcal{U}$ (corresponding to a single application) transmits a vector of packets $\mathbf{x}_{u}(t)$ to the \gls{bs}.
All packets have the same length $L$, and we denote by $x_{u}(t, m)$ the $m$-th packet of $\mathbf{x}_{u}(t)$.
We also assume that each application is associated with a specific slice $\sigma \in \Sigma$, according to the application requirements.
We denote by $\mathcal{U}_{\sigma} \subset \mathcal{U}$ the set of users associated with slice $\sigma$.
In particular, all the packets belonging to the same slice $\sigma$ are seen by the \gls{bs} as a single stream of data sharing the same communication resources.

We assume that the \gls{bs} maintains a \gls{fifo} buffer with a maximum size of $Q$ packets for each slice $\sigma \in \Sigma$. The packets present in the buffer for slice $\sigma$ at the beginning of slot $t$ are collected in vector $\mathbf{q}_{\sigma}(t)$. We can conservatively assume that packets are added to the backhaul link buffer at the end of each slot. The buffer size condition is then $|\mathbf{q}_{\sigma}(t)| \leq Q$,
where $|\mathbf{x}|$ represents the length of vector $\mathbf{x}$. 

We assume that the backhaul link has a total capacity $C_{\text{bh}}$, that can be divided into $N$ resource blocks, each of which makes it possible to transmit $\frac{\tau C_{\text{bh}}}{N}$ bits per slot.
Note that packets can be transmitted over multiple subsequent slots.
We now denote by $N_{\sigma}(t)$ the number of resource blocks assigned to slice $\sigma$ during slot $t$.
Naturally, any resource allocation scheme should comply with the condition  $\sum_{\sigma\in\Sigma}N_{\sigma}(t) \leq N$.

The number of packets from slice $\sigma$ that can be delivered in slot $t$, given that $N_{\sigma}(t)$ backhaul resources are allocated to it, is then given by:
\begin{equation}
  \chi_{\sigma}(t)=\min\left(|\mathbf{q}_{\sigma}(t)|, \left\lfloor  N_{\sigma}(t) \frac{\tau C_{\text{bh}}}{LN}\right\rfloor \right).
\end{equation}
We denote by $\mathbf{y}_{\sigma}(t)$ the vector containing the $\chi_{\sigma}(t)$ packets of slice $\sigma$ that are transmitted during slot $t$.
At the next step, the queue vector $\mathbf{q}_{\sigma}(t+1)$ contains the remaining queued packets, as well as the newly arrived ones. However, the queue cannot contain more than $Q$ packets, and in case of buffer overflows, the oldest $\omega_{\sigma}(t)$ packets are then discarded:
\begin{equation}
  \omega_{\sigma}(t)=\max\left(0,|\mathbf{q}_{\sigma}(t-1)|-\chi_{\sigma}(t-1)+|\mathbf{x}_{\sigma}(t)|-Q\right).
\end{equation}
We denote the vector of discarded packets by $\mathbf{d}_{\sigma}(t)$, so that $|\mathbf{d}_{\sigma}(t)| = \omega_{\sigma}(t)$. Once the buffer states $\mathbf{q}_{\sigma}(t),\,\forall\sigma \in \Sigma$ have been updated, the backhaul resources can be reallocated. To evaluate the system performance, we consider the queuing delay, defined as:
\begin{equation}
  \delta(p,t)=\begin{cases}
              s:p\in\mathbf{x}_{\sigma}(t-s), &\text{ } p\in\left(\mathbf{y}_{\sigma}(t)\cup\mathbf{q}_{\sigma}(t)\right);\\
              \infty, &\text{ }  p\in\mathbf{d}_{\sigma}(t),
            \end{cases}
\end{equation}
where the elay is computed for packet $p$ from slice $\sigma$ at time $t$.
Therefore, the delay is the number of slots since $p$ was delivered to the \gls{bs} if $p$ is transmitted or still in the buffer, while it is considered infinite if $p$ is discarded.

We then define a utility function $f_{u}(\cdot)$ for each user, which takes the packet delay as input and return a value in $[0,1]$.
In particular, $f_{u}(\cdot)$ is monotonically decreasing, with $f_u(0)=1$ and $\lim_{x\rightarrow\infty}f_{u}(x)=0$, and depends on the delay requirements of the user, expressed in terms of the maximum delay $\Delta_u$.
Finally, the performance $\Phi_u(t)$ of $u$ and the overall system performance $\Phi(t)$ at slot $t$ are given by:
\begin{equation}
\Phi_u(t)=\sum_{m=1}^{\chi_{\sigma}(t)}\frac{f_{u(y_{\sigma}(t,m))}(\delta(y_{\sigma}(t,m),t))}{(\chi_{\sigma}(t)+\omega_{\sigma}(t))},
\label{eq:reward_user}
\end{equation}
\begin{equation}
\Phi(t)=\frac{1}{|\Sigma|}\sum_{\sigma\in\Sigma}\frac{1}{|\mathcal{U}_{\sigma}|}\sum_{u\in\mathcal{U}_{\sigma}}\Phi_u(t),\label{eq:reward}
\end{equation}
where $u(p)$ is the user that sent packet $p$.

\subsection{Learning framework}
\label{sec:learn-model}

In order to optimize the system in a foresighted manner, we model the resource allocation problem as an \gls{mdp}, implementing a \gls{drl} agent to manage the link resources. 
In particular, the state $s(t)$ at time $t$ depends on the individual requirements of the users currently being served, as well as the state of the transmission buffer for each slice. Specifically, $s(t)$ is a tuple with $4\cdot|\Sigma| $ elements, namely:

\begin{itemize}
	\item The number of packets contained in each slice buffer, i.e., $|\mathbf{q}_{\sigma}(t)|$, $\forall$ $\sigma \in \Sigma$;
	\item The average remaining time before the packets contained in each slice buffer exceed the maximum allowed delay $\Delta_u$, which is given by	
	\begin{equation}
		\sum_{m=1}^{|\mathbf{q}_{\sigma}(t)|}\frac{\Delta_{u(q_{\sigma}(t,m))}-\delta(q_{\sigma}(t,m))}{|\mathbf{q}_{\sigma}(t)|},\ \forall\sigma \in \Sigma;
	\end{equation} 
	\item The minimum remaining time among the packets in each slice buffer, i.e.,
	\begin{equation}
	 \min_{m\in\{1,\ldots,|\mathbf{q}_{\sigma}(t)|\}} \left(\Delta_{u(q_{\sigma}(t,m))}-\delta(q_{\sigma}(t,m))\right),\ \forall\sigma \in \Sigma;
	\end{equation}
	\item The number of packets that will be transmitted during the current slot for each slice, assuming that resource allocation scheme does not change, which is:
	\begin{equation}
		\min\left(|\mathbf{q}_{\sigma}(t)|, \left\lfloor  N_{\sigma}(t-1) \frac{\tau C_{\text{bh}}}{LN}\right\rfloor \right),\ \forall \sigma \in \Sigma.
	\end{equation}
\end{itemize}
We have $\mathcal{S}=\left\{0,\ldots,\frac{Q}{L}\right\}^{2\cdot|\Sigma|}\times\mathbb{R}^{2\cdot|\Sigma|}$, while the action space $\mathcal{A}$ includes $1+|\Sigma|\cdot(|\Sigma|-1)$ different actions.
Specifically, action 0 maintains the resource allocation constant, so that $N_{\sigma}(t)=N_{\sigma}(t-1)$, $\forall$ $\sigma \in \Sigma$. 
Instead, each of the remaining actions is defined by the ordered tuple $(i,j),\,i\neq j$, and corresponds to taking one resource block from slice $\sigma_i$ and assigning it to slice $\sigma_j$, so that $N_{\sigma_i}(t)=N_{\sigma_i}(t-1)-1$ and $N_{\sigma_j}(t)=N_{\sigma_j}(t-1)+1$.
Naturally, resource allocation can never be negative, i.e., $N_{\sigma}(t)\geq0\ \forall\sigma\in\Sigma$, and the total number of allocated resources is always $N$.

Hence, at each slot $t$, the agent observes state $s(t) \in \mathcal{S}$ and selects a new action $a(t) \in \mathcal{A}$ according to its current policy $\pi$, which determines the number of blocks $N_{\sigma}(t)$ assigned to each slice $\sigma\in\Sigma$.
Then, the agent receives an instantaneous reward $\Phi(t)$, which is given by the system utility defined in~\eqref{eq:reward}; we note that, by definition, $\Phi(t)\in[0,1]$.

We adopt the \gls{dqn} approach~\cite{mnih2015human}, in which the expected long-term value of each action, as given by~\eqref{eq:long_term} with $\Phi(t)$ as reward, is approximated by a \gls{nn} that, at each slot $t$, takes the current state $s(t)$ as input.
In particular, we implement a fully connected feed-forward \gls{nn}, whose input layer is formed by one neuron for each element of $s(t)$. 
The output of the \gls{nn} is a scalar vector of size $|\mathcal{A}|$, representing the expected long-term reward of each possible action $a \in \mathcal{A}$ for the current state.
In particular, when operating greedily, the agent will always pick the action corresponding to the highest output value.
The main parameters of the learning architecture are reported in Tab.~\ref{tab:learn_param}.

\begin{table}[t!]
	\vspace{.3cm}
	\centering
		\scriptsize
	\caption{Agent architecture.}
	\label{tab:learn_param}
	\begin{tabular}{cc}
		
		\toprule
		Layer size (inputs $\times$ outputs) & Inter-layer operations \\
		\midrule
		
		 $4 \cdot |\Sigma|\times64$ & ReLU activation \\
		
		$64\times32$ & ReLU activation \\
		
		$32\times(1 + |\Sigma|\cdot(|\Sigma|-1))$ & Linear activation \\
		
		\bottomrule
	\end{tabular}
	\vspace{-0.25cm}
\end{table}

\section{Settings and Results}
\label{sec:results}

In the following, we apply our meta-\gls{rl} model in the \gls{ns} environment described in Sec.~\ref{sec:system-model}. 
Specifically, we assume that the \gls{drl} agent deployed at the \gls{bs} cannot be trained locally and needs to exchange information with the core network to improve its own policy.  
Part of the backhaul link resources are then used to update the agent's architecture, possibly degrading the user performance.
Hence, we investigate how the system utility changes when varying the resources dedicated to the training, both when the learning update is performed regularly in time, and when it is stopped after a certain period (assuming the \gls{drl} agent has converged to the optimal policy). 
In the rest of the section, we will describe the settings we used in our system, as well as the simulation results.

\subsection{Scenario settings}

The model described in Sec.~\ref{sec:comm-model} is very general and can suit multiple communication scenarios with different characteristics. 
In this work, for the sake of simplicity, we consider a simple case with only two slices, named \emph{non-critical} ($\sigma_{\text{NC}}$) and \emph{critical} ($\sigma_{\text{C}}$), respectively, with the following performance functions $f_u(\delta)$:
\begin{align}
f_{u} (\delta) =\begin{cases}
\min\left(1,\frac{\Delta_u}{\delta}\right), &\text{if } u\in\sigma_{\text{NC}}; \\
\mathbbm{1}(\Delta_u-\delta), &\text{if } u\in\sigma_{\text{C}};
\end{cases}
\end{align}
where $\mathbbm{1}(x)$ is the limit-step function, equal to 1 if $x\geq0$ and 0 otherwise.
Critical packets have a hard deadline, i.e., delivering them after the maximum delay $\Delta_u$ gives zero performance benefits.
On the other hand, the utility of non-critical packets decreases gradually as the delay grows past the deadline.

We assume that time is divided into slots of $\tau=10$~ms and that the backhaul link has a total capacity $C_{\text{bh}}=1$~Mb/s.
The channel is divided into $N=10$ resource blocks so that each block allows the delivery of exactly 1~kb.
However, we assume that any allocation scheme remains constant over 10 slots, i.e., the agent takes a new action every 100~ms.
This value is closer to the granularity of actual schedulers and avoids the \emph{reward tampering} phenomenon~\cite{everitt2021reward}.
Finally, we consider packets to have a constant length $L=512$ b, and set the maximum buffer size to $Q = 100$ packets, equivalent to 64 kB.

We assume that there are 5 different users in the system, each of which randomly picks an application among the 2 critical and 2 non-critical applications at the beginning of each episode.
The \gls{ncvo} and \gls{ncvi} applications are associated with $\sigma_{\text{NC}}$ and have constant bitrate.
The \gls{cvo} and \gls{cvi} applications are associated with $\sigma_{\text{C}}$ and generate new packets following an on-off process. 

\begin{table}[t!]
	\vspace{.3cm}
	\centering
	\scriptsize
	\caption{Application parameters.}
	\label{tab:app_param}
	\begin{tabular}{l|ll}
		\toprule
		Application & 
		Bit rate [kb/s] & Packet delay budget [ms] \\
		\midrule
		\gls{ncvo} & 25 & 100 \\
		\gls{ncvi} & 384 & 300 \\
		\gls{cvo} & 25 (when active) & 75 \\
		\gls{cvi} & 384 (when active) & 100 \\
		
		\bottomrule
	\end{tabular}
\vspace{-.25cm}
\end{table}

More specifically, \gls{cvo} and \gls{cvi} behave according to a \gls{mc} with two state, namely \emph{silent} ($s$) and \emph{active} ($a$): in the silent states, no packets are generated, which implies that $|\mathbf{x}_{u}(t)|= 0$; conversely, in the active states, packets are generate with constant bitrate. 
We assume that a critical application can switch between states at the beginning of each slot $t$, and we set the state transition probability to $p_{ss}=p_{aa}=0.9$ and $p_{sa}=p_{as}=0.1$, respectively. 
The bitrate and the packet delay budget of the different applications are summarized in Tab.~\ref{tab:app_param}.

\subsection{Learning settings}

To orchestrate the backhaul link resources, we deploy the learning system presented in Sec.~\ref{sec:learn-model}.
Our implementation of \gls{dqn} is distributed: the \emph{inference network} is installed on the \gls{bs} and is used to allocate resources in the exploitation phase of each episode, while the \emph{training network} is updated at every training step and is in a Cloud server.
The transmission of the learning data and updated model occupies the backhaul link, following the general framework defined in Sec.~\ref{sec:meta-model}. 

At the beginning of each episode, the slice buffers are emptied and each user is associated with a random application. 
During the first $T_{\pi}$ slots of each episode, the inference network sets the resource allocation policy and saves experience samples $\left(s(t), a(t), s(t+1), \Phi(t), a(t+1)\right)$ in a local memory. 
Instead, during the training phase, no application data are transmitted, i.e., $N_{\sigma}(t)=0$, $\forall$ $\sigma \in \Sigma$, and the entire link capacity is used to forward training data to the core network.

We assume that each experience sample can be encoded into $L_{\text{tr}}=704$ b, while the \gls{nn} architecture size is $L_{\text{NN}}=92256$ b.
Therefore, during the updating phase, $ T_{\rho} \frac{\tau C_{\text{bh} }} {L_{\text{tr}}}$ transitions can be forwarded through the channel and used by the Cloud server to update the training network.
Moreover, every $10$ learning steps, a copy of the training network is sent back to the \gls{bs}, replacing the inference network and improving the current policy $\rho$. 
In those steps, the number of transitions forwarded through the channel is limited to $ \frac{T_{\rho} \tau C_{\text{bh} } - L_{\text{NN}}} {L_{\text{tr}}}$. 

To train the agent during the update phase, we implement the \emph{on-policy} \gls{sarsa} algorithm~\cite{sutton2018reinforcement} with a 
\emph{softmax} exploration policy. 
In particular, we set the discount factor $\gamma=0.95$, and we implement the \gls{adam} algorithm to optimize the \gls{nn} weights, using $\zeta=10^{-5}$ as maximum learning rate.
Finally, we set the coherence period to $K=10000$ episodes, and we assume that each episode lasts $T=1000$ slots. 

\subsection{Results}

\begin{figure}
	\centering
	% This file was created with tikzplotlib v0.9.15.
\begin{tikzpicture}

\pgfplotsset{every tick label/.append style={font=\scriptsize}}

\begin{axis}[
width=\fwidth,
height=\fheight,
legend cell align={left},
ylabel style={font=\footnotesize\color{white!15!black}},
xlabel style={font=\footnotesize\color{white!15!black}},
legend style={font=\tiny, at={(0.99,0.02)}, legend columns=2, anchor=south east, legend cell align=left, align=left,fill opacity=0.8, draw opacity=1, text opacity=1, draw=white!80!black},
tick align=outside,
tick pos=left,
x grid style={white!69.0196078431373!black},
xlabel={Episode},
xmajorgrids,
xmin=196, xmax=9803,
xtick style={color=black},
y grid style={white!69.0196078431373!black},
ylabel={\(\displaystyle \Phi\)},
ymajorgrids,
ymin=0.775, ymax=0.875,
ytick style={color=black}
]
\addplot [line width=0.4pt, color0, mark=+, mark size=1.5, mark options={solid}]
table {%
196 0.825923085212708
392 0.851661384105682
588 0.853278577327728
784 0.855794608592987
980 0.859286248683929
1176 0.860682606697083
1372 0.862481355667114
1568 0.860346436500549
1764 0.859801590442657
1960 0.860648274421692
2156 0.86005038022995
2352 0.861229479312897
2549 0.861341416835785
2745 0.86319488286972
2941 0.862651348114014
3137 0.862483739852905
3333 0.861186325550079
3529 0.862798273563385
3725 0.86229133605957
3921 0.861782252788544
4117 0.862319231033325
4313 0.86234313249588
4509 0.862054169178009
4705 0.862919807434082
4901 0.861272990703583
5098 0.862577676773071
5294 0.864006221294403
5490 0.86389148235321
5686 0.862958192825317
5882 0.864095270633698
6078 0.862598478794098
6274 0.864316403865814
6470 0.862531244754791
6666 0.861535847187042
6862 0.863844633102417
7058 0.861427664756775
7254 0.865016639232635
7450 0.864975214004517
7647 0.86375105381012
7843 0.861285328865051
8039 0.862913191318512
8235 0.864414036273956
8431 0.863240480422974
8627 0.861532092094421
8823 0.862749993801117
9019 0.863412141799927
9215 0.864204406738281
9411 0.863134443759918
9607 0.862831056118011
9803 0.864195942878723
};
\addlegendentry{Ideal}
\addplot [line width=0.4pt, color1, mark=x, mark size=1.5, mark options={solid}]
table {%
196 0.769278705120087
392 0.769878745079041
588 0.773737847805023
784 0.775773167610168
980 0.779166162014008
1176 0.780453383922577
1372 0.782089948654175
1568 0.785135567188263
1764 0.787788569927216
1960 0.790180444717407
2156 0.791951298713684
2352 0.791471600532532
2549 0.796485483646393
2745 0.795930087566376
2941 0.800486743450165
3137 0.800051629543304
3333 0.804447293281555
3529 0.804506778717041
3725 0.805756151676178
3921 0.804410457611084
4117 0.808814644813538
4313 0.810703277587891
4509 0.810584425926208
4705 0.813348829746246
4901 0.814054489135742
5098 0.816299021244049
5294 0.817217290401459
5490 0.816908240318298
5686 0.816972970962524
5882 0.818282902240753
6078 0.819337844848633
6274 0.822886228561401
6470 0.821094632148743
6666 0.822793543338776
6862 0.824920296669006
7058 0.823606252670288
7254 0.826542794704437
7450 0.827419936656952
7647 0.828003227710724
7843 0.82786762714386
8039 0.829302430152893
8235 0.828569233417511
8431 0.828553140163422
8627 0.831820487976074
8823 0.832364022731781
9019 0.832185328006744
9215 0.83204460144043
9411 0.832399010658264
9607 0.83392745256424
9803 0.834516823291779
};
\addlegendentry{$T_{\rho} = 1$}
\addplot [line width=0.4pt, color2, mark=diamond, mark size=1.5, mark options={solid}]
table {%
196 0.766718924045563
392 0.771154165267944
588 0.775267779827118
784 0.779607892036438
980 0.784455537796021
1176 0.788894891738892
1372 0.793594896793365
1568 0.796294927597046
1764 0.800954163074493
1960 0.803807318210602
2156 0.807882845401764
2352 0.810429036617279
2549 0.811231017112732
2745 0.8151975274086
2941 0.816716134548187
3137 0.81752347946167
3333 0.821428000926971
3529 0.822482466697693
3725 0.823759019374847
3921 0.826886475086212
4117 0.826912939548492
4313 0.82613468170166
4509 0.828418850898743
4705 0.829901814460754
4901 0.830598890781403
5098 0.83220237493515
5294 0.831435263156891
5490 0.831705152988434
5686 0.834015667438507
5882 0.834659039974213
6078 0.836056590080261
6274 0.835955202579498
6470 0.838283360004425
6666 0.837219178676605
6862 0.838159322738647
7058 0.839219689369202
7254 0.836581110954285
7450 0.838654339313507
7647 0.838891744613647
7843 0.837689161300659
8039 0.840405583381653
8235 0.840884745121002
8431 0.84222012758255
8627 0.841282725334167
8823 0.84262603521347
9019 0.84113883972168
9215 0.842098653316498
9411 0.840343415737152
9607 0.841879427433014
9803 0.841967940330505
};
\addlegendentry{$T_{\rho} = 2$}
\addplot [line width=0.4pt, color3, mark=triangle, mark size=1.5, mark options={solid,rotate=180}]
table {%
196 0.758369207382202
392 0.766084790229797
588 0.775276243686676
784 0.781108140945435
980 0.787082672119141
1176 0.792559504508972
1372 0.799904704093933
1568 0.804323077201843
1764 0.807481706142426
1960 0.812461972236633
2156 0.814425766468048
2352 0.819840669631958
2549 0.820777952671051
2745 0.822703957557678
2941 0.826558411121368
3137 0.828716516494751
3333 0.830121099948883
3529 0.831656396389008
3725 0.833926141262054
3921 0.834186732769012
4117 0.83632355928421
4313 0.83730810880661
4509 0.836901426315308
4705 0.838483929634094
4901 0.83741295337677
5098 0.838416218757629
5294 0.84171199798584
5490 0.83816260099411
5686 0.840376138687134
5882 0.840971231460571
6078 0.840028166770935
6274 0.840341866016388
6470 0.841329038143158
6666 0.841893792152405
6862 0.840893685817719
7058 0.842485725879669
7254 0.842990875244141
7450 0.842248916625977
7647 0.838530838489532
7843 0.840387880802155
8039 0.843580722808838
8235 0.842613577842712
8431 0.84358811378479
8627 0.841657817363739
8823 0.842453300952911
9019 0.841228842735291
9215 0.841215670108795
9411 0.84331077337265
9607 0.843464434146881
9803 0.843073844909668
};
\addlegendentry{$T_{\rho} = 3$}
\addplot [line width=0.4pt, color4, mark=o, mark size=1.5, mark options={solid}]
table {%
196 0.752224981784821
392 0.763073742389679
588 0.773580253124237
784 0.783875644207001
980 0.790059030056
1176 0.797029137611389
1372 0.804305136203766
1568 0.808062136173248
1764 0.813465416431427
1960 0.815292179584503
2156 0.817859172821045
2352 0.819272220134735
2549 0.822191059589386
2745 0.824114441871643
2941 0.825859904289246
3137 0.828521251678467
3333 0.829778373241425
3529 0.829157292842865
3725 0.832122385501862
3921 0.83223569393158
4117 0.833339631557465
4313 0.834457039833069
4509 0.834710657596588
4705 0.834260523319244
4901 0.834615111351013
5098 0.8351029753685
5294 0.835812985897064
5490 0.837558031082153
5686 0.835565984249115
5882 0.836458146572113
6078 0.836181163787842
6274 0.836294889450073
6470 0.83791595697403
6666 0.836686909198761
6862 0.838845729827881
7058 0.838127136230469
7254 0.836300253868103
7450 0.837325692176819
7647 0.836439192295074
7843 0.836706936359406
8039 0.837580442428589
8235 0.837500333786011
8431 0.837060034275055
8627 0.837123572826385
8823 0.837359249591827
9019 0.837347865104675
9215 0.836759090423584
9411 0.836285889148712
9607 0.83700168132782
9803 0.837725341320038
};
\addlegendentry{$T_{\rho} = 4$}  
\addplot [line width=0.4pt, color5, mark=Mercedes star flipped, mark size=1.5, mark options={solid}]
table {%
196 0.764224052429199
392 0.775722920894623
588 0.78629195690155
784 0.792145550251007
980 0.800758481025696
1176 0.805984020233154
1372 0.809108912944794
1568 0.81329482793808
1764 0.816707134246826
1960 0.82000207901001
2156 0.821498095989227
2352 0.824445784091949
2549 0.824208915233612
2745 0.824620306491852
2941 0.827974319458008
3137 0.826032876968384
3333 0.827063143253326
3529 0.827616631984711
3725 0.829579830169678
3921 0.829957664012909
4117 0.829268336296082
4313 0.830271482467651
4509 0.829787909984589
4705 0.832260489463806
4901 0.830784916877747
5098 0.830797672271729
5294 0.832134068012238
5490 0.833620429039001
5686 0.832854807376862
5882 0.830678701400757
6078 0.834316372871399
6274 0.832671344280243
6470 0.833011746406555
6666 0.831257283687592
6862 0.833217144012451
7058 0.831377685070038
7254 0.831518530845642
7450 0.832286655902863
7647 0.831412374973297
7843 0.829982280731201
8039 0.83101898431778
8235 0.830886662006378
8431 0.830725848674774
8627 0.830950260162354
8823 0.832562685012817
9019 0.831238269805908
9215 0.830789625644684
9411 0.832309424877167
9607 0.833279609680176
9803 0.832578480243683
};
\addlegendentry{$T_{\rho} = 5$}
\end{axis}

\end{tikzpicture}
	\caption{Mean performance over time with fixed $T_{\rho}$.}
	\label{fig:linear_performance}
\end{figure}
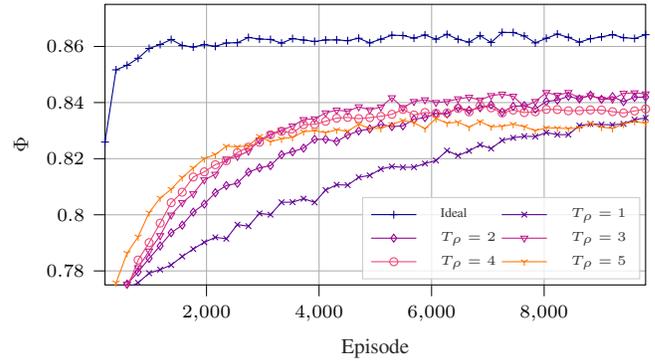

We now investigate different configurations of our learning system, varying the time $T_{\rho}\in\{1,\ldots,5\}$ devoted to the agent training. 
When $T_{\rho}$ is low, most of the system resources are assigned to the users, and agent training proceeds slowly; conversely, as $T_{\rho}$ increases, more learning data are transmitted through the link, taking up more resources but increasing the training speed. 
We compare the results with an ideal system, where all the learning transitions are instantaneously transmitted through a side-channel and used to update the inference network, without impacting the users. Clearly, this represents an upper bound to the practically achievable performance. 

We first consider the scheme with a constant update duration from Sec.~\ref{ssec:const}. In Fig.~\ref{fig:linear_performance} we represent the average performance during the training of the different strategies, obtained aggregating the data of 100 independent simulations.
It is easy to see that larger values of $T_{\rho}$ lead to a quicker convergence toward the optimal policy and, in particular, training is rather slow if $T_{\rho}=1$.
On the other hand, the configurations with a higher $T_{\rho}$ improve faster but have lower performance after convergence, since they continue to devote a fixed amount of resources to the agent training. 

\begin{figure}
	\centering
	\input{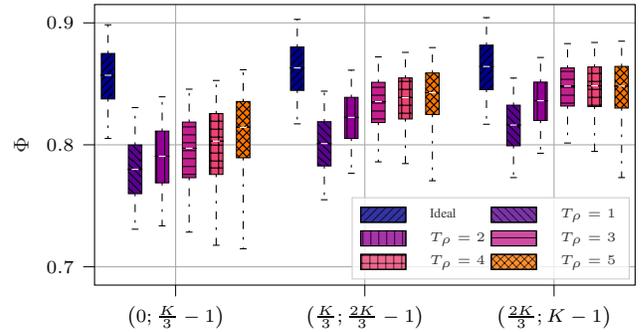}
	\caption{Boxplots of the performance with fixed $T_{\rho}$.}
	\label{fig:box_performance}
	\vspace{-0.4cm}
\end{figure}

We can have a better insight into the different strategies by looking at Fig.~\ref{fig:box_performance}, which uses boxplots to represent the performance statistics during the beginning, intermediate and last episodes of the training. The whiskers represent the $5$th and $95$th percentiles of the performance distribution, while the box goes from the $25$th to the $75$th. 
In the first third of the coherence period, it is convenient to select a high value for $T_{\rho}$ to learn the optimal policy faster. 
After a sufficient number of episodes, the configurations with a lower $T_{\rho}$ also converge and waste fewer system resources on further training, but the configuration with $T_{\rho}=1$ is always outperformed by the others, since it needs more than $K$ episodes to reach convergence.

We can now look at the adaptive system from Sec.~\ref{ssec:adaptive}.
We use a simple heuristic strategy to infer convergence: if the average reward over a rolling window of $K_{\text{avg}}$ episodes stops increasing, the agent estimates that the optimal policy has been found and sets $T_{\rho}$ to zero, assigning the whole link capacity to the users. 
This allows the agent to use the first portion of the coherence period to learn the optimal policy, taking advantage of the acquired knowledge in the remaining episodes. The choice of $T_{\rho}$ will then determine the time needed to switch to the full exploitation phase.

\begin{figure}
	\centering
	% This file was created with tikzplotlib v0.9.15.
\begin{tikzpicture}

\pgfplotsset{every tick label/.append style={font=\scriptsize}}

\begin{axis}[
width=\fwidth,
height=\fheight,
legend cell align={left},
ylabel style={font=\footnotesize\color{white!15!black}},
xlabel style={font=\footnotesize\color{white!15!black}},
legend style={font=\tiny, at={(0.99,0.02)}, legend columns=2, anchor=south east, legend cell align=left, align=left,fill opacity=0.8, draw opacity=1, text opacity=1, draw=white!80!black},
tick align=outside,
tick pos=left,
x grid style={white!69.0196078431373!black},
xlabel={Episode},
xmajorgrids,
xmin=196, xmax=9803,
xtick style={color=black},
y grid style={white!69.0196078431373!black},
ylabel={\(\displaystyle \Phi\)},
ymajorgrids,
ymin=0.775, ymax=0.875,
ytick style={color=black}
]
\addplot [line width=0.4pt, color0, mark=+, mark size=1.5, mark options={solid}]
table {%
196 0.825923085212708
392 0.851661384105682
588 0.853278577327728
784 0.855794608592987
980 0.859286248683929
1176 0.860682606697083
1372 0.862481355667114
1568 0.860346436500549
1764 0.859801590442657
1960 0.860648274421692
2156 0.86005038022995
2352 0.861229479312897
2549 0.861341416835785
2745 0.86319488286972
2941 0.862651348114014
3137 0.862483739852905
3333 0.861186325550079
3529 0.862798273563385
3725 0.86229133605957
3921 0.861782252788544
4117 0.862319231033325
4313 0.86234313249588
4509 0.862054169178009
4705 0.862919807434082
4901 0.861272990703583
5098 0.862577676773071
5294 0.864006221294403
5490 0.86389148235321
5686 0.862958192825317
5882 0.864095270633698
6078 0.862598478794098
6274 0.864316403865814
6470 0.862531244754791
6666 0.861535847187042
6862 0.863844633102417
7058 0.861427664756775
7254 0.865016639232635
7450 0.864975214004517
7647 0.86375105381012
7843 0.861285328865051
8039 0.862913191318512
8235 0.864414036273956
8431 0.863240480422974
8627 0.861532092094421
8823 0.862749993801117
9019 0.863412141799927
9215 0.864204406738281
9411 0.863134443759918
9607 0.862831056118011
9803 0.864195942878723
};
\addlegendentry{Ideal}
\addplot [line width=0.4pt, color1, mark=x, mark size=1.5, mark options={solid}]
table {%
196 0.770766973495483
392 0.773186147212982
588 0.777280271053314
784 0.779053747653961
980 0.781893372535706
1176 0.784851491451263
1372 0.785784780979156
1568 0.787329614162445
1764 0.791231393814087
1960 0.792915821075439
2156 0.797045767307281
2352 0.797137975692749
2549 0.79766720533371
2745 0.80128401517868
2941 0.801326513290405
3137 0.804761826992035
3333 0.807020366191864
3529 0.806862652301788
3725 0.808880686759949
3921 0.811708092689514
4117 0.811722099781036
4313 0.812725722789764
4509 0.815564155578613
4705 0.817503869533539
4901 0.818301618099213
5098 0.818218231201172
5294 0.820936143398285
5490 0.820842981338501
5686 0.821074962615967
5882 0.822194218635559
6078 0.823801457881927
6274 0.824368715286255
6470 0.825385093688965
6666 0.826939821243286
6862 0.827616155147552
7058 0.828815579414368
7254 0.82690954208374
7450 0.829471945762634
7647 0.829399943351746
7843 0.830686867237091
8039 0.831178843975067
8235 0.834503710269928
8431 0.831017732620239
8627 0.831837892532349
8823 0.833973288536072
9019 0.83324408531189
9215 0.833289742469788
9411 0.836101293563843
9607 0.832992553710938
9803 0.836159110069275
};
\addlegendentry{$T_{\rho} = 1$}
\addplot [line width=0.4pt, color2, mark=diamond, mark size=1.5, mark options={solid}]
table {%
196 0.756580650806427
392 0.757624506950378
588 0.765861809253693
784 0.76975017786026
980 0.775129437446594
1176 0.780638456344604
1372 0.786816596984863
1568 0.787616968154907
1764 0.793061137199402
1960 0.799030005931854
2156 0.800353705883026
2352 0.804675757884979
2549 0.807200133800507
2745 0.80961811542511
2941 0.812115132808685
3137 0.813084602355957
3333 0.816397428512573
3529 0.818328559398651
3725 0.818798542022705
3921 0.821128964424133
4117 0.82262521982193
4313 0.825588047504425
4509 0.826641619205475
4705 0.827572703361511
4901 0.827443420886993
5098 0.829549849033356
5294 0.830528140068054
5490 0.832659780979156
5686 0.83284866809845
5882 0.83394181728363
6078 0.834568440914154
6274 0.836258471012115
6470 0.837037622928619
6666 0.838311433792114
6862 0.837389409542084
7058 0.839172661304474
7254 0.838478565216064
7450 0.840005397796631
7647 0.838069260120392
7843 0.842020928859711
8039 0.840648055076599
8235 0.843376159667969
8431 0.842648029327393
8627 0.843280375003815
8823 0.840854465961456
9019 0.844169139862061
9215 0.842834293842316
9411 0.84484601020813
9607 0.845218002796173
9803 0.844254314899445
};
\addlegendentry{$T_{\rho} = 2$}
\addplot [line width=0.4pt, color3, mark=triangle, mark size=1.5, mark options={solid,rotate=180}]
table {%
196 0.758905589580536
392 0.764822900295258
588 0.772355318069458
784 0.779080510139465
980 0.7867711186409
1176 0.790916383266449
1372 0.797090291976929
1568 0.800925314426422
1764 0.804183602333069
1960 0.8079514503479
2156 0.810827970504761
2352 0.814165472984314
2549 0.815616428852081
2745 0.817324817180634
2941 0.819635987281799
3137 0.821866393089294
3333 0.823379993438721
3529 0.825432956218719
3725 0.826888680458069
3921 0.828156709671021
4117 0.827885091304779
4313 0.830329358577728
4509 0.830350458621979
4705 0.83251816034317
4901 0.832606017589569
5098 0.834430396556854
5294 0.83382910490036
5490 0.835388541221619
5686 0.835723698139191
5882 0.836058795452118
6078 0.836502015590668
6274 0.836998641490936
6470 0.838058352470398
6666 0.839894652366638
6862 0.839442312717438
7058 0.840876340866089
7254 0.841398894786835
7450 0.843155682086945
7647 0.84080570936203
7843 0.842303276062012
8039 0.84269905090332
8235 0.842594027519226
8431 0.843910157680511
8627 0.84471070766449
8823 0.844136416912079
9019 0.843478739261627
9215 0.845566749572754
9411 0.845775663852692
9607 0.845788836479187
9803 0.844727575778961
};
\addlegendentry{$T_{\rho} = 3$}
\addplot [line width=0.4pt, color4, mark=o, mark size=1.5, mark options={solid}]
table {%
196 0.770703375339508
392 0.780610620975494
588 0.788158655166626
784 0.795046925544739
980 0.799775958061218
1176 0.805967926979065
1372 0.80890953540802
1568 0.8144850730896
1764 0.818306505680084
1960 0.818437993526459
2156 0.821302592754364
2352 0.824134826660156
2549 0.825882792472839
2745 0.826888561248779
2941 0.829485714435577
3137 0.828171789646149
3333 0.828584909439087
3529 0.830853462219238
3725 0.831178426742554
3921 0.834088683128357
4117 0.836606621742249
4313 0.837528467178345
4509 0.836837291717529
4705 0.837208569049835
4901 0.839865267276764
5098 0.841048717498779
5294 0.840236723423004
5490 0.841180622577667
5686 0.841649413108826
5882 0.841839134693146
6078 0.841616690158844
6274 0.843632519245148
6470 0.843628764152527
6666 0.843870341777802
6862 0.845732688903809
7058 0.845753610134125
7254 0.846264958381653
7450 0.846653401851654
7647 0.847799241542816
7843 0.847840249538422
8039 0.847023069858551
8235 0.847171187400818
8431 0.848764061927795
8627 0.848895311355591
8823 0.849485874176025
9019 0.851057231426239
9215 0.851203382015228
9411 0.850039482116699
9607 0.851373612880707
9803 0.85153603553772
};
\addlegendentry{$T_{\rho} = 4$}
\addplot [line width=0.4pt, color5, mark=Mercedes star flipped, mark size=1.5, mark options={solid}]
table {%
196 0.759179592132568
392 0.769191026687622
588 0.780408978462219
784 0.790548622608185
980 0.795632481575012
1176 0.802448689937592
1372 0.803992629051208
1568 0.808798670768738
1764 0.811473608016968
1960 0.813347101211548
2156 0.815460503101349
2352 0.816472232341766
2549 0.819538712501526
2745 0.8212611079216
2941 0.823310911655426
3137 0.822443664073944
3333 0.825511574745178
3529 0.826407432556152
3725 0.825464963912964
3921 0.827837407588959
4117 0.830360949039459
4313 0.831161797046661
4509 0.831128299236298
4705 0.833620488643646
4901 0.833140313625336
5098 0.83438378572464
5294 0.837153375148773
5490 0.835123836994171
5686 0.838342547416687
5882 0.838184297084808
6078 0.839690446853638
6274 0.839757561683655
6470 0.840025067329407
6666 0.840542793273926
6862 0.840473115444183
7058 0.840233027935028
7254 0.84336906671524
7450 0.842808306217194
7647 0.843465566635132
7843 0.843192636966705
8039 0.843504846096039
8235 0.844619274139404
8431 0.845242917537689
8627 0.847791910171509
8823 0.846979796886444
9019 0.847191095352173
9215 0.846787929534912
9411 0.847574055194855
9607 0.848213672637939
9803 0.84732723236084
};
\addlegendentry{$T_{\rho} = 5$}
\end{axis}

\end{tikzpicture}
	\caption{Mean performance over time with adaptive $T_{\rho}$.}
	\label{fig:linear_performance_adaptive}
\end{figure}
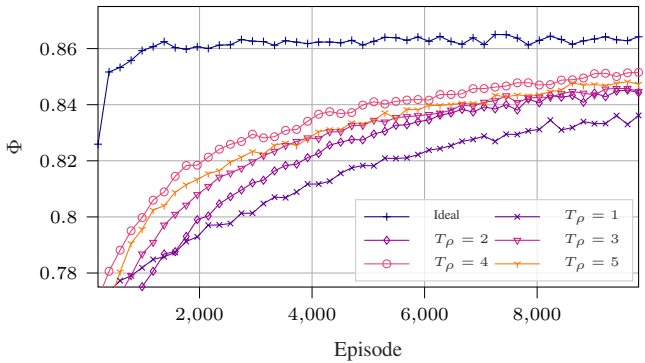

\begin{figure}
	\centering
	\input{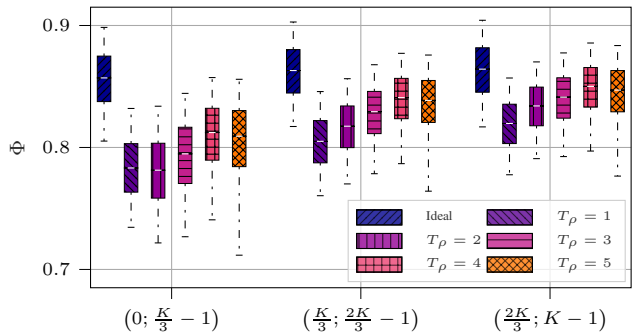}
	\caption{Boxplots of the performance with adaptive $T_{\rho}$.}
	\label{fig:box_performance_adaptive}
	\vspace{-0.4cm}
\end{figure}

Fig.~\ref{fig:linear_performance_adaptive} shows the performance over time with this approach, considering $K_{\text{avg}}=4000$.
We observe that the performance of the strategies with $T_{\rho} \leq 3$ is almost identical to that of the previous scenario.
In fact, the adaptive method does not reach convergence before the end of the coherence period with such a small $T_{\rho}$.
On the other hand, setting $T_{\rho} = 5$ seems to be too aggressive, and the heuristic stops the learning process too soon, leading to suboptimal results. 

The approach with $T_{\rho}=4$ outperforms all the other strategies during the entire coherence period, striking a balance between convergence speed and cost of learning.
This is confirmed by Fig.~\ref{fig:box_performance_adaptive}, which shows that $T_{\rho}=4$ achieves better performance, both considering the lower and the higher percentiles of the distribution.

\section{Conclusions}
\label{sec:concl}

In this work, we analyzed the cost of exploiting \gls{drl} solutions for \gls{mec} network optimization. 
Specifically, we designed a novel cost of learning framework to optimize the amount of resources that a learning agent allocates to its own improvement, so as to balance the speed of convergence of the policy with the system performance during the training.
We consider a test case based on a 5G system in which a \gls{drl} agent has to allocate the bandwidth of a backhaul link among multiple slices while consuming part of the network resources to transmit its own training updates.
Our results show that there is a significant trade-off between the convergence speed and the communication overhead due to the training.
Inferring the agent convergence is also a critical problem, especially in fast-varying scenarios which require the use of continual learning.
Future work on the subject may involve more general solutions, based on hierarchical \gls{drl} or other meta-learning tools, that can apply to different problems in which cost of learning is a real concern.

\bibliographystyle{IEEEtran}
\bibliography{bibliography} 

\end{document}